\documentclass[twocolumn,showpacs,preprintnumbers]{revtex4}
\usepackage{graphicx}
\usepackage{dcolumn}
\usepackage{bm}
\usepackage{color}
\usepackage{amsmath}
\usepackage{amsfonts}
\usepackage[pdftex]{epsfig}
\usepackage{epstopdf}
\usepackage{float}
\usepackage{subfigure}
\usepackage{caption}

\begin{document}

\title[]{Superconducting properties of $FeSe_{0.5}Te_{0.5}$ under high pressure}
\author{ Esmeralda Mart\'inez-Pi\~neiro$^1$}
\author{Roberto Escudero}
\affiliation{Instituto de Investigaciones en Materiales, Universidad Nacional Aut\'{o}noma de M\'{e}xico. A. Postal 70-360. M\'{e}xico, D.F., 04510 M\'EXICO}
\email[Author to whom correspondence should be addressed. Dr. Roberto Escudero. email address:] {escu@unam.mx}

\date{\today}

\begin{abstract}
This work reports studies performed in the superconducting compound $FeSe_{0.5}Te_{0.5}$ under high pressure. Changes were observed  in the transition temperature, superconducting critical fields, anomalous variations in the Meissner fraction, and in  Ginzburg Landau parameters. The superconducting properties were calculated and compared using the Werthamer-Helfand-Hohenber approximation and  Ginzburg-Landau theory.  Hydrostatic  pressure was produced  from atmospheric   to   823 MPa, the increment   in  the critical temperature was from 14.45 to 20.5 K  at  a  rate of change about   0.0069 K/MPa.
\end{abstract}


\maketitle
\section{Introduction}
The recent discovery  of the Fe-based superconductors  has proved that superconductivity and ferromagnetism may  coexists with the  consequent changes  on
the electronic pair formation,  different to the electron-phonon type described by  the BCS theory \cite{Kordyuk}. The new type  of superconducting materials opens  other  possibilities for  the electron pairing and  the  perspectives to explore  different processes on  superconducting compounds  with magnetic elements, for instances the recent results  by  Nakayama et al., in  $FeSe$ film with K doping, discovered  that the cross-interfacial electron-phonon coupling was  not the primary interaction for  the superconducting pairing \cite{Nakayama, Mazin}.
The superconducting family of compounds  with Fe and chalcogenides, i.e.,   $Fe(Se,Te)$  has a     simple crystal structure   known  as the anti-PbO,  with $P4/mmm$ symmetry \cite{KW}. Physical modifications on these structures, particularly on  the  Fe-Se bonds give place to other superconducting compounds    with different transition temperatures  \cite{Yeh}. Yeh, et al., have investigated that  the  transition temperature in these   compounds changes because  the coupling   between layers can be modified  via   chemical and/or external  pressures \cite{Fong, Stems, Yoshi, Mizugu}. One  example of these   changes was observed in the $FeSe$ \cite{BC}  where the  transition temperature $T_C$ changes with the internal pressure when  selenium is substituted by  tellurium  resulting the family of $FeSe_{1-x}Te_x$.  In the particular case of $FeSe_{0.5}Te_{0.5}$ the internal pressure  increases  the $T_C$ from  8  to  15 K \cite{Horigante}. Moreover, other manner  to produce changes in the electronic and structural properties on these compounds is by applying   high pressures by  external means. The  tools  to produce external pressures are called pressure cells. The most powerful of these instruments are the diamond pressure cells that are able to produce pressures of the order of  hundred of GPa at hydrostatic or quasi-hydrostatics forms. Here in this work we used a   pressure cell without diamonds that is able to produce in a simple manner hydrostatic pressures at  the order of a few MPa. So, this cell was used  to investigate  the range not yet  studied, of the variations of critical fields and Ginzburg-Landau parameters at pressures of  only a  few MPa.

Recent pressure studies  of  the $Fe(Se,Te)$  family  were  focused  in the  observation of changes  on $T_C$ and in its crystal structure. In a  report by Stemshorn, et al. \cite{stemshorn}, the authors indicate that the $FeSe$  tetragonal structure ($P4/nmm$) was transformed to  an  hexagonal one ($P6_{3}/mmc$)  with  applied pressure and becoming amorphous at higher pressures. On the other hand, it was observed that  $FeSe_{0.5}Te_{0.5}$ increased its $T_C$ to 26 K  at $P=2$ GPa while higher pressures produces structural transitions to a monoclinic symmetry that decreases $T_C$ \cite{horigane,Tsoi,Huang}. Moreover at $P=11.5$ GPa  the compound changes  into an   amorphous phase \cite{gresty,Pallavi}. In a recent report Pietosa et al.,  studied the  pressure effects on  the  critical fields up to  1.04 GPa  founding  no important variations on the behavior of the   $dH_{C2}/dT$, concluding  that  the observed changes on the upper critical field   was related only to  changes of   $T_C$  \cite{Pietosa}.

In this work we  investigated  the behavior and changes  on the  critical fields of $FeSe_{0.5}Te_{0.5}$ under hydrostatic pressures below 900 MPa.  Our investigation pointed out  that  this behavior   is almost  independent of the variations  of  $T_C$
The   primary objective  of this research  is to study the behavior on the critical fields, Ginzburg-Landau parameters and thermodynamic properties at  different pressures.  It is important  to mention that  this superconductor   has been scarcely  studied under the influence of  external pressures below 1 GPa.  Other researchers  have attributed those  changes to variations   on the density of the superconducting carriers. However,  that   analyses by Fedor, et al.,  was not very conclusive and the main part of their investigations  was mainly  focused to observe the  $T_C$ variations instead on the behavior superconducting properties \cite{Fedor}.

\section{Experimental details}
$FeSe_{0.5}Te_{0.5}$ samples were prepared by  solid state reactions starting with stoichiometric amounts of selenium powder (Alfa Aesar, $99.99\%$), tellurium powder
 (Alfa Aesar, $99.999\%$) and iron pieces (Merck, $99.999\%$), were mixed and  powdered in an agatha mortar and  pelletized. The resulting samples were vacuum-sealed in quartz tubes and heated at  $1000 ^{o}C$ during 40 h. After this procedure were  cooled at a rate of $10^{o} C/h$ following the   similar procedure as already published  \cite{Velasco,Awana}. Crystal structural
   analysis was performed by X-ray diffraction and scanning electron microscopy. X-ray studies  were obtained using a D5000-Siemens diffractometer with $Co-K\alpha$ radiation.
    Compositions of the samples were estimated using a field emission scanning electron microscopy (JEOL JSM 7600F) coupled with energy dispersive x-ray spectroscopy (EDX, Oxford).
Superconducting properties were determined by magnetic measurements with  a SQUID based magnetometer (Quantum Design MPMS). High pressure measurements were performed with a CuBe cell (Quantum Design)
 using a small piece of Pb as the manometer, for the  pressure media was used Daphne 7373 oil. In order to subtract the magnetic background the cell was  measured without sample  \cite{QD}. Daphne pressure oil medium was used because of its lack of magnetic contribution when the temperature decreases \cite{Kamara}. The superconducting transition temperatures were determined  by  magnetic susceptibility with  field of  20 Oe oriented  parallel to the ab-plane in two modes, zero field cooling  ZFC, and field cooling FC.

\section{Results and discussion}
 Diffraction analysis determined that the  compound has  $FeSe_{0.44}Te_{0.56}$ with very   small  traces of $\beta -FeSe$, as  shown in Fig 1. The   compound  presents  anti-PbO crystalline  structure and  $P4/nmm$ symmetry (CSD 421334) \cite{Tegel}  with  preferential  orientation on the plane [100], the non superconducting compound $\beta-FeSe$ has NiAs-type structure and $P6_{3}/mmc$ symmetry (ICDD 85-0735)  \cite{Gomez}.

A  small sample  was studied with SEM, it  shows   lamellar structure, see    Fig 1 inset,  confirming that the compound was highly oriented.
Backscattered electron images reveals regions of  light gray homogeneous coloration corresponding to richer parts of  Te  as already explained by Pimentel et al., \cite{Pimentel}, while dark zones are produced by the surface morphology. EDX analysis was used to determine  that  averaged composition was  $Fe_{1.09}Se_{0.45}Te_{0.55}$,  results were consistent with SEM images and DRX analysis.

\begin{figure}[ht]
\includegraphics[scale=1.3]{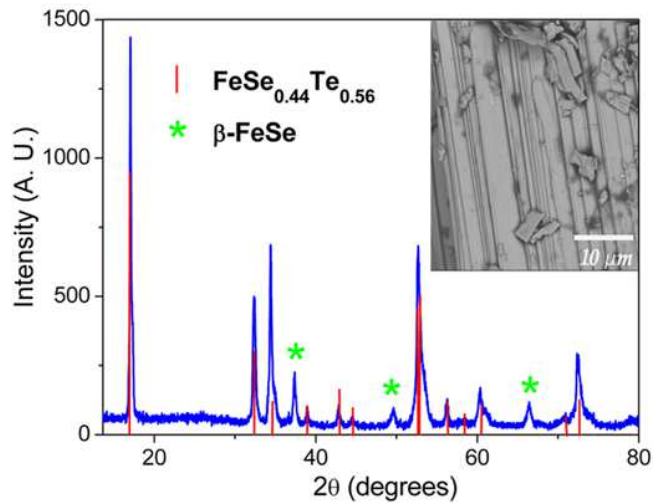}
\caption{(Color online) Diffraction patterns of  sample, $FeSe_{0.44}Te_{0.56}$ with small traces of the non-superconducting FeSe Achavalite structure, those   reflections are signaled  with $*$. The inset shows SEM images of backscattering electrons and the  homogeneous layered sample.}
\label{fig1}
\end{figure}

 Superconducting properties  of $Fe_{1.09}Se_{0.45}Te_{0.55}$ were   studied at low temperatures and ambient pressure with  DC Magnetic susceptibility, $\chi(T)$. Those measurements   are presented in  Fig. 2. The inset shows a plot from 2 o 30  K.   Demagnetization factor was estimated to be 0.826 due to  the shape  and size of the sample \cite{Amikam,Dajero}. Fig. 2 also shows a  small paramagnetic contribution due to $\beta-FeSe$, this is seen above the superconducting transition in the inset.  There, also  is shown the ZFC mode which is  about 50 times bigger than the FC mode. It is important to mention that this behavior has  been attributed  to strong pinning effects as in \cite{Bendele}.

\begin{figure}[h]
\centerline{\includegraphics[scale=1.7]{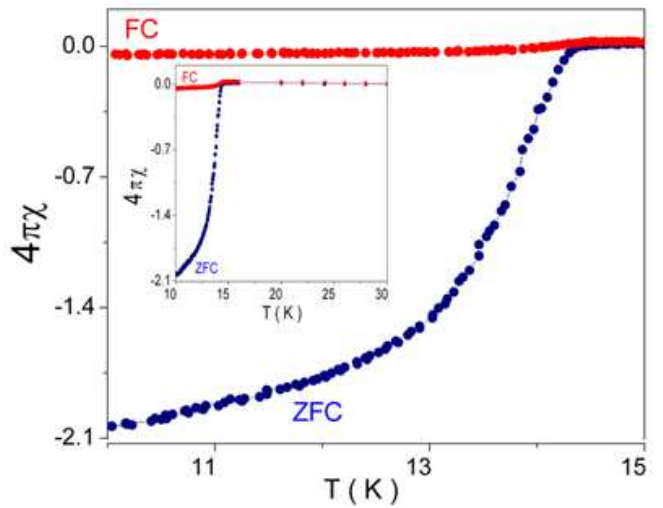}}\caption{(Color online)   Transition temperature, $T_C = 14$ K of the compound $FeSe_{0.44}Te_{0.56}$ measured in FC and ZFC modes. The inset  shows the temperature dependence of the DC magnetic susceptibility at low temperatures.}
\label{fig2}
\end{figure}

 Once the sample  was characterized at room atmospheric pressure, high  pressure measurements were performed  with the sample inside of  the  CuBe  cell. DC magnetic measurements were performed in two  modes  under different hydrostatic pressures. The results of these measurements are  illustrated  in Fig. 3, measurements are displayed in terms  of $4\pi \chi $ units.

\begin{figure}
\includegraphics[scale=1.6]{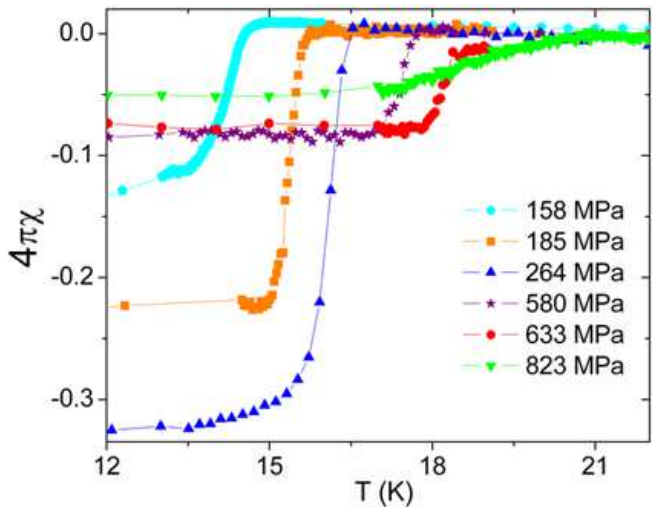}
\caption{(Color online)Field cooling measurements near $T_C$ at different pressures.}
\label{fig3}
\end{figure}

The   transition temperature was taken when  $\chi(T)$ curve deviates from the zero background value. The results show  that  $T_C$ increases continuously with the pressure.  Fig. 4   shows a plot of the changes experimented of the $T_C$   by pressure effects with  the linear fit of $T_C - P$  extracted from Fig 3.

\begin{figure}[h]
\centerline{\includegraphics[scale=1.4]{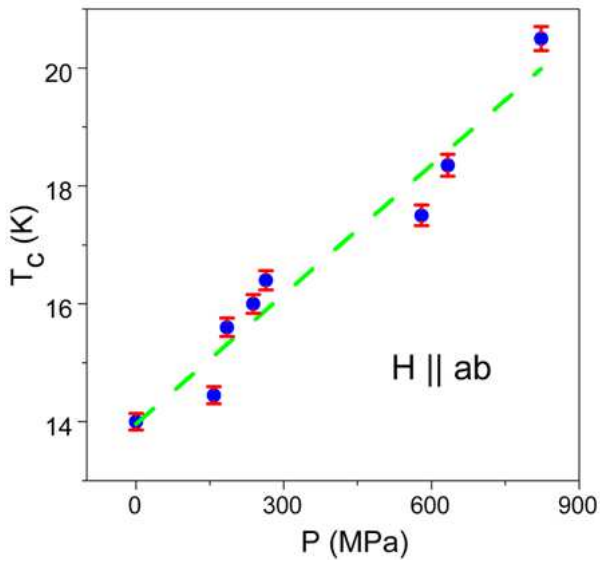}}
\caption{(Color online) Changes experimented by  the transition temperature with pressure, $T_C- P$. The rate of change is linear and varies as  0.0069 K/MPa similar to other measurements. }
\label{fig4}
\end{figure}

  The Meissner fraction was calculated with   $-4\pi\chi=-4\pi\rho M/{mB}$, where $\rho$ is the sample density, equal to $6.06 g/cm^{3}$, $M$ the magnetization, $m$  the  mass of the sample  and $B$  the applied magnetic field \cite{Lai}. Fig. 5 shows  the Meissner fraction  determined at different pressures. The Meissner fraction  increases with pressure from $13\%$ at a $P=158$ MPa to  about $34\%$ at $P=264$ MPa. With higher pressures, about $P=580$ MPa  the fraction decreases at about  $8\%$.  We have to note that an anomalous increase  is observed at  264  MPa,  the fraction is highly increased, and  with increasing pressure it decreases dramatically. In other studies  in $BaFe_{2}(As_{1-x}P_{x})_{2}$ \cite{Shuai} and  $La_{2-x}Sr_{x}CuO_{4}$ \cite{Zhou} it was observed a similar behavior in experiments with substituting atoms and producing  internal pressures. Those  similar experiments demonstrate   that  the Meissner fraction could varies in a non-monotonic way with the pressure. However we believe that more experimental work will  be  necessary in order to investigate more about this behavior. Lastly, it is  very important to mention that   in most of the literature never was studied  or  observed the  Meissner fraction changes  because  the  signals  were very small   and, the  FC and ZFC measurements have an enormous differences between them.

\begin{figure}
\includegraphics[scale=1.3]{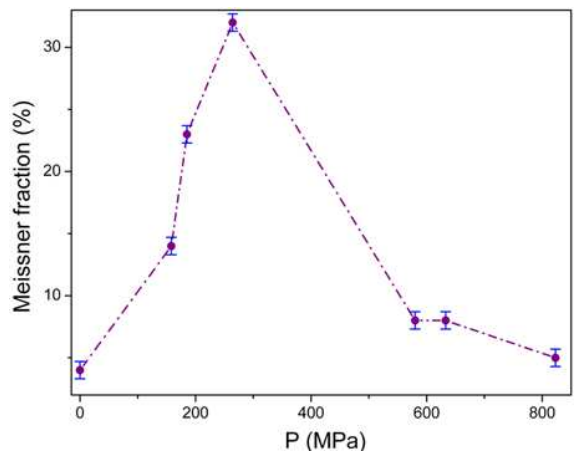}
\caption{(Color online) Meissner fraction changes  with pressure in the $Fe_{1.09}Se_{0.45}Te_{0.55}$ sample.}
\label{fig5}
\end{figure}

Critical magnetic  field measurements were performed and determined with  isothermal magnetization curves. For determinations of  the lower critical fields, $H_{C1}$ we used in those curves the point where the diamagnetic curve start  to  deviate  from  linearity. $H_{C1}(T)$  temperature dependence  was  fitted with equation:   $H_{C1}(T)=H_{C1}(0)(1-(T/T_{C})^2)$, as shown in fig. 6 and  in Table 1.

\begin{figure}[ht]
\centerline{\includegraphics[scale=1.3]{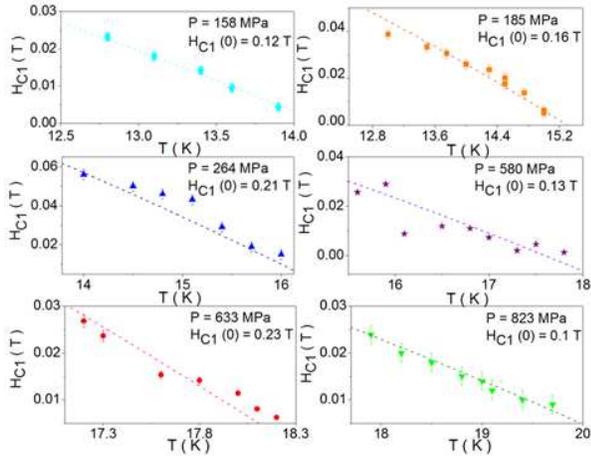}}
\caption{(Color online) Lower critical field, $H_{C1}(0)$ versus applied pressure  estimated with the formula described in the text.}
\label{fig6}
\end{figure}

The upper critical field $H_{C2}$ was calculated with  the linear fit from experimental data near $T_{C}$, and using the  Werthamer-Helfand-Hohenberg (WHH) approximation \cite{Li}. This approximation is given by;   $H_{C2}(0)=-0.693T_C\frac{dH_{C2}(T)}{dT} \mid_{T=T_{C}}$, where  $\frac{dH_{C2}(T)}{dT} \mid_{T=T_{C}}$ corresponds to the slope of the linear fit. The  Ginzburg-Landau parameters,  coherence length $\xi_{GL}$, penetration length $\lambda_{GL}$, $\kappa$, and the thermodynamic critical field $H_{C}(0)$, were estimated with  the equations:
$H_{C2}(0)=\phi_{0}/2\pi\xi_{GL}^{2}$,
$H_{C2}(0)/H_{C1}(0)=2\kappa^{2}/\ln\kappa$,
$\kappa=\lambda_{GL}/\xi_{GL}$,
$H_{C}(0)=\phi_{0}/(2\sqrt{2}\pi\xi_{GL}\lambda_{GL})$,
where $\phi{_0}$ is the quantum magnetic flux.

In this analysis the upper critical field shows  a concave upward feature near  $T_C$,  see  illustration  in  Fig 7. This behavior is attributed to multiband response and has been also observed  in other iron-selenides compounds by Jing, et al., and Bezusyy, et al., \cite{Jing, Bezu} and in $Na_{0.35}CoO_{2}yH_{2}O$ by Kao, et al.,\cite{Kao}. For our calculations of  the critical fields, we use a linear fit excluding the curve zone, as it was performed  by Pietosa. Note that the slope $-dH_{C}/dT$ has different values for each pressure. In some investigations \cite{Pietosa}, the authors have considered that the increment on the critical fields is caused only by the increment   of $T_C$. Our experiments clearly indicate that the critical fields reach the highest value at $P=600$  MPa, instead of 823 MPa, these values are shown in fig. 8. In order to assure that this behavior is correct, our measurements were repeated at least twice using different samples.

\begin{figure}[ht]
\centerline{\includegraphics[scale=1.3]{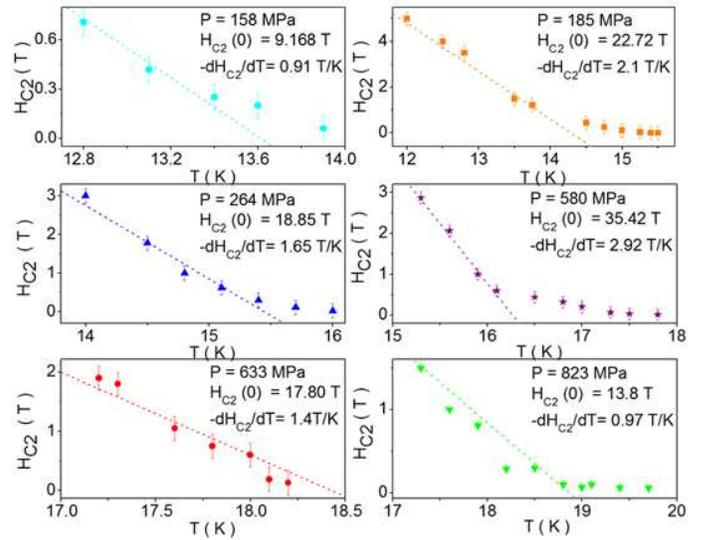}}
\caption{{(Color online)Upper critical fields versus  temperature  determined with  the WHH aproximation.} }
\label{fig7}
\end{figure}

\begin{table*}[h]
\caption{\label{superparam} Superconducting parameters}
\small\rm
\begin{tabular}{c c c c c c c c c}
\hline
P	&	$T_{C}$ 	&	$H_{C1}(0)$	&	-$dH_{C2}/dT$	&	$H_{C2}(0)$	&	$\xi_{GL}$	&		$\kappa$	&	$H_{C} (0)$	& $\lambda_{GL}$	\\
(MPa)	&	(K)	&	(T)	&	(T/K)	&	(T)	&	(nm)	&		&(T)		&	(nm)	\\
\hline
0	&	14.5	&	0.03(0)	&	0.15(3)	&	1.50(7)	&	1.4(7)	&	6.9(5)	&	0.15(2)	&	102.69(7)	\\
158	&	14.7	&	0.12(2)	&	0.91(2)	&	9.16(8)	&	5.9(9)	&	9.2(0)	&	0.70(3)	&	55.1(2)	\\
185	&	15.6	&	0.16(0)	&	2.10(2)	&	22.72(4)	&	3.8(1)	&	13.6(1)	&	1.18(1)	&	51.7(5)	\\
264	&	16.4	&	0.21(3)	&	1.65(9)	&	18.85(7)	&	4.1(8)	&	10.2(90)	&	1.30(4)	&	42.6(1)	\\
580	&	17.5	&	0.13(4)	&	2.92(1)	&	35.428(4)	&	3.0(5)	&	19.6(1)	&	1.25(7)	&	59.7(4)	\\
633	&	18.4	&	0.23(3)	&	1.40(2)	&	17.80(3)	&	4.2(9)	&	9.2(5)	&	1.35(2)	&	39.7(7)	\\
823	&	20.5	&	0.10(1)	&	0.97(0)	&	13.78(0)	&	4.8(8)	&	13.3(1)	&	0.72(7)	&	65.0(5)	\\

\hline
\end{tabular}
\end{table*}

\begin{figure}[ht]
\centerline{\includegraphics[scale=1.3]{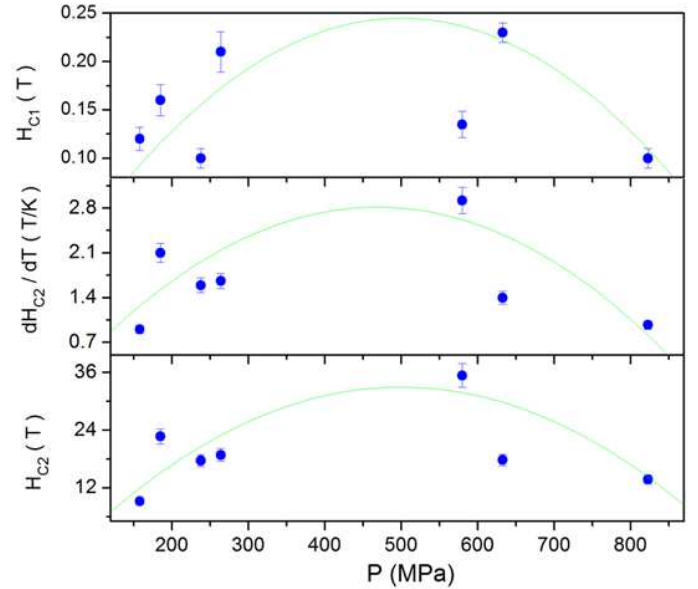}}
\caption{{(Color online)Critical fields versus  pressure.} }
\label{fig8}
\end{figure}

With the help of  Ginzburg-Landau theory the superconducting parameters  were determined and  shown  in Table 1. Those parameters  are similar to   data  measured at normal ambient pressure  \cite{Horigante,Velasco,Awana}.   We found that some  parameters varies in irregular form and present anomalous values. $\lambda_{GL}$ for instances  decreases more than $50\%$ as pressure is applied, and then shows a small increment at higher pressure. This may implies a considerable increment on the superconducting electronic density. Coherence length also  is increased by about  $400\%$  at $P=158$ MPa. According to our experiments, and others,  pressure strongly affect the superconducting parameters and these changes can be  attributed not only to  increment of $T_{C}$ but to additional factors, as crystal parameters and structural changes, etc \cite{Pallavi}. In our investigation, we noticed that the $T_{C}$, critical fields and superconducting parameter are very sensitive to pressure, being a clear sign that this compound presents unconventional superconductivity \cite{minev}.

\section{Conclusions}

Magnetic studies under hydrostatic pressure  were performed in $FeSe_{0.5}Te_{0.5}$ in order  to investigate the impact of pressure on the basic parameters of this  superconductor. We found a linear increase of  $T_{C}$  that changes from 14.03 K to 20.5 K,  at a rate of 0.0069 K/MPa. The anomalous behavior of the Meissner fraction and superconducting parameters are difficult to explain because those are not directly related to  changes of the transition temperature with pressure. As mentioner earlier, the changes in pressure may be  associated with $Se-Fe-Se$ interlayer separation  \cite{Margadonna}. On the other hand, the dependence of superconducting properties with pressure of this compound implying that microscopic mechanism of the electronic pairing is different of electron-phonon, as it is shown in unconventional superconductors.  Other reports on  similar studies \cite{Lu, Ciechan}, indicate that the increment of the density of electronic states are not enough to  explanation the notable   increment  of the transition temperature. The presence of a Peierls distortion, as  a  Spin Density Wave,  must  be decreased by pressure but nevertheless has  any influence on  the transition temperature, again this is quite anomalous. Lastly, we must mention that still more experiments are necessary in order to completely understand this type of new superconducting materials \cite{koufos,Mandal}.

\begin{acknowledgements}
EM acknowledge financial support of a scholarship given by CONACYT, and PCeIM-UNAM. Also we thanks  to  DGAPA-UNAM project IN106014.
\end{acknowledgements}

\end{document}